\documentstyle[12pt,citesort]{article}  

\newcommand{\gsim}{\mathrel{\mathop{\kern 0pt \rlap
  {\raise.2ex\hbox{$>$}}}
  \lower.9ex\hbox{\kern-.190em $\sim$}}}  

\begin{document}
\begin{titlepage}
%\begin{flushright}
%UAB--FT--???\\
%June 2003
%\end{flushright}
\vspace*{1.cm}

\begin{center}
{\Large\bf Quantitative duality and neutral kaon interferometry in CPLEAR
experiments}\\
\vspace*{0.8cm}

A. Bramon$^{1}$, G. Garbarino$^{2}$ and B.~C.~Hiesmayr$^{1}$\\
\vspace*{0.2cm}

{\footnotesize\it
$^1$Grup de F{\'\i}sica Te\`orica,
Universitat Aut\`onoma de Barcelona, \\
E--08193 Bellaterra, Spain\\
$^2$Departament d'Estructura i Constituents de la Mat\`{e}ria,
Universitat de Barcelona, \\ E--08028 Barcelona, Spain}
\end{center}
\vspace*{3.0cm}

\begin{abstract}
A quantitative formulation of Bohr's complementarity principle and interferometric 
duality is discussed and applied to the neutral kaon system. Recent
measurements by the CPLEAR Collaboration can be easily interpreted in
terms of ``\emph{neutral kaon interferometry}'' illustrating and confirming those basic
principles of quantum mechanics.     
\end{abstract}

\vspace*{1.2cm}

\vspace*{1.0mm}  

\end{titlepage}
%%%%%%%%%%%%%%%%
\date{\today}
%%%%%%%%%%%%%%%
\section{Introduction}
%{\bf 1.- Introduction}
 
In his well known ``Lectures on Physics'', Feynman starts discussing the
double--slit experiment as the most  characteristic feature of quantum
mechanics \cite{feynman}:      
\begin{quotation} 
In reality, it contains the \emph{only} mistery. 
\end{quotation}
After this frequently quoted statement, an idealized but detailed analysis of the 
double--slit interference phenomenona is presented in terms
of wave--particle duality, the rules for superposition of amplitudes and
Bohr's complementarity  principle.   Somewhat later, in the same ``Lectures'',
Feynman proceeds to  discuss a particularly illustrative case: the neutral
kaon system, for which he drastically concludes that \cite{feynman}:
\begin{quotation} 
If there is any place where we have a chance to test the main 
principles of quantum mechanics in the purest way ---does the superposition of
amplitudes  work or doesn't it?--- this is it.
\end{quotation}
In the present letter we reconsider these issues with a twofold
purpose in mind. 
First, we are going to show that a quantitative statement on duality ---or
``interferometric duality'', as recently suggested by Englert
\cite{englert96} and reviewed in 
Refs.~\cite{englert99,englert00,durr00}---, which was originally
proposed by Greenberger and Yasin \cite{GY88}, can be clearly illustrated using
the $K^0$--$\bar{K^0}$ system.  Secondly, this will then allow us to
interpret two CPLEAR experiments on neutral kaons performed at CERN
\cite{CPLEARlepton,CPLEARstrong,CPLEARreview}
as quantitative and elegant tests of
duality and Bohr's complementarity in a new arena, which we would
like to refer as ``neutral kaon interferometry''.
 
%\vspace{0.5cm}
%{\bf 2.- Quantitative duality}
\section{Quantitative duality} 

The quantitative  expression for Bohr's complementarity    
proposed by Greenberger and Yasin \cite{GY88} is extremely simple. 
In today's most common notation \cite{englert96,englert99,englert00,durr00}, it reads:  
\begin{equation}
\label{PV}
{\cal P}^2 + {\cal V}_{0}^2 \le 1, 
\end {equation}
where ${\cal V}_{0}$ is the ``fringe visibility'' which quantifies the sharpness 
or contrast of the interference pattern (a wave--like 
property), whereas  ${\cal P}$ is the path ``predictability''  
quantifying the \emph{a priori} ``which--way'' knowledge one has on 
the path taken by the interfering system (its complementary 
particle--like property).  
In commonly used two--path interferometers ---such as single--crystals in
neutron interferometry \cite{GY88,RS84} or their Mach--Zehnder analogs
in optical experiments, e.g. in Ref.~\cite{MPS87}--- one has to deal with
two--level quantum  states of the type 
\begin{equation}
\label{state}
|\Psi (\phi )\rangle =a |\psi_{I}\rangle + b\, e^{i \phi}|\psi_{II}\rangle, 
\end {equation}
where $a$ and $b$ are positive, $a^2 + b^2 =1$ and $|\psi_{I}\rangle$ and 
$|\psi_{II}\rangle$ represent the states corresponding 
to the two spatially separated interferometric paths, with (ideally) 
$\langle \psi_{II}|\psi_{I}\rangle=0$  and a controllable relative phase $\phi$. 
In the case of symmetric interferometers ($a =b =1/\sqrt{2}$), 
the two paths are taken with the same  probability, thus no \emph{a priori} ``which--way''
knowledge is available, ${\cal P} =0$, 
and maximal ``fringe visibility'' is expected, ${\cal V}_0 =1$. 
In asymmetric cases ($a \ne b$),
instead, the  expression (\ref{PV}) becomes more interesting and quantifies
the  simultaneous wave and particle knowledge one can have 
for the interfering object according to Bohr's complementarity. 
Note that we are refering to \emph{a priori} knowledge, which is fully
contained in the preparation of the state (\ref{state}). We thus exclude any 
possibility of knowledge--improving measurements (otherwise, another
interesting inequality has been derived by Englert \cite{englert96} and
reviewed in Refs.~\cite{englert99,englert00,durr00}; its applicability to
entangled neutral kaon systems has been recently proposed in Ref.~\cite{BGH03}). 
If no knowledge--improving measurement 
is performed, state (\ref{state}) remains \emph{pure}
and fully coherent, and expression (\ref{PV}) is then verified with the 
\emph{equal} sign. 

The proof of this equality is quite simple, once the correct
definitions and meanings for ${\cal V}_0$ and ${\cal P}$ are identified.  The ``fringe
visibility'', ${\cal V}_{0}$, is defined in the usual way as
the coefficient ${\cal V}_{0} \equiv (I_{max} - I_{min})/ (I_{max} + I_{min})$
of the phase--dependent term in the expressions:
\begin{equation}
\label{I}
I(\phi){_{\pm}} = \left|\langle\psi_{\pm}|\Psi (\phi )\rangle \right|^2 = 
{1 \over 2} [ 1 \pm {\cal V}_{0} \cos \phi ] ,
\end {equation}
which give two output intensities in two--channel
interferometers in terms of $\phi$.  
The amplitudes corresponding to paths $|\psi_I\rangle$ and $|\psi_{II}\rangle$ 
are obtained by a first splitting of the initial beam and
are later recombined before emerging from the two
output ports. Most frequently and in Eq.~(\ref{I}), these two outputs are
associated to measurements on the symmetric and antisymmetric basis states:  
\begin{equation}
\label{pm}
|\psi_{\pm}\rangle  = {1 \over \sqrt 2} [|\psi_{I}\rangle  \pm |\psi_{II}\rangle ].
\end {equation}
For the coherent superposition (\ref{state}), rewritten in this new basis:  
\begin{equation}
\label{statepm}
|\Psi (\phi )\rangle = {1 \over \sqrt 2} (a + b\, e^{i \phi}) |\psi_{+}\rangle +
{1 \over \sqrt 2} (a - b\, e^{i \phi}) |\psi_{-}\rangle ],
\end {equation}
one easily obtains:
\begin{equation}
\label{V}
{\cal V}_{0} = 2ab . 
\end{equation}
The crucial concept of path ``predictability'', ${\cal P}$, turned out to be much harder to
identify and was first introduced in Ref.~\cite{GY88}. Denoting with
$w_I$ and $w_{II}$ the probabilities for taking either one of the two interferometric
paths, ${\cal P}$ is defined as  
\begin{equation}
\label{P}
{\cal P} \equiv \left|w_{I} - w_{II}\right| = \left|a^2 - b^2\right| ,
\end {equation}
where the final equality is specific of our pure state (\ref{state}). From 
Eqs.~(\ref{V}) and (\ref{P}), the desired equality:
\begin{equation}
\label{PV=}
{\cal P}^2 + {\cal V}_{0}^2 = 1 ,
\end {equation}
follows immediately. 
The use of the path ``predictability'' ${\cal P}$ introduced in Ref.~\cite{GY88}
and defined in Eq.~(\ref{P}) ---as opposed to other measures of
``which--way'' knowledge, as those introduced in Ref.~\cite{mandel91} or in the 
theoretical--information approaches of Refs.~\cite{RS84,MPS87,WZ79} (for
a critical discussion,  see Ref.~\cite{JSV95})--- has been crucial in order to derive
Eq.~(\ref{PV=}).   This equality can then be viewed as a modern and
quantitative statement of Bohr's complementatity principle for pure states
like (\ref{state}). 

%\vspace{0.5cm}
%{\bf 3.- Neutral kaon system}
\section{Neutral kaon system}

Pure states of neutral kaons have been copiously prepared at CPLEAR 
\cite{CPLEARlepton,CPLEARstrong,CPLEARreview} using proton--antiproton
annihilations at rest, $p \bar p \to K^- \pi^+ K^0$ or 
$p \bar p \to K^+ \pi^- \bar{K^0}$, where strangeness conservation in
strong interactions requires that a $K^-$ has to be accompanied by a $K^0$
and a $K^+$ by a $\bar{K^0}$. In free space, the initially produced 
${K^0}$ and $\bar{K^0}$ evolve in proper time $\tau$ according to the well
known expressions \cite{kabir}:
\begin{eqnarray}
\label{statet}
|K^0 (\tau)\rangle  &=& {\sqrt{1+|\epsilon |^2} \over \sqrt{2} (1 + \epsilon )} 
\left[Êe^{-i\lambda_S \tau} |K_S\rangle  + e^{-i\lambda_L \tau} |K_L\rangle \right] , \nonumber \\
|\bar{K^0} (\tau)\rangle  &=& {\sqrt{1+|\epsilon |^2} \over \sqrt{2} (1 - \epsilon )} 
\left[Êe^{-i \lambda_S \tau} |K_S\rangle  - e^{-i \lambda_L \tau} |K_L\rangle  \right], \nonumber 
\end{eqnarray}
where $\epsilon$ is the (small) $CP$--violation parameter, 
$\lambda_{S,L} \equiv m_{S,L} -{i \over 2}\Gamma_{S,L}$ and   
$m_{S,L}$ and $\Gamma_{S,L}$ are the masses and decay widths of the 
short-- or long--lived states, $K_S$ or $K_L$. 
By normalizing to kaons surviving up to time $\tau$, the previous states can be
more conveniently rewritten as:
\begin{eqnarray}
\label{statetn}
|K^0 (\tau)\rangle  &=& {1 \over \sqrt{1 + e^{-\Delta \Gamma \tau}}} 
\left[ |K_S\rangle  + e^{-{1\over 2} \Delta \Gamma \tau}
e^{-i \Delta m \tau} |K_L\rangle \right] , \\ 
|\bar{K^0} (\tau)\rangle  &=& {1 \over \sqrt{1 + e^{-\Delta \Gamma \tau}}} 
\left[ |K_S\rangle  - e^{-{1\over 2} \Delta \Gamma \tau}
e^{-i \Delta m \tau} |K_L\rangle \right], \nonumber
\end{eqnarray}
where $\Delta m \equiv m_L -m_S$, $\Delta \Gamma \equiv \Gamma_L - \Gamma_S$ 
and, even if one strictly has
$\langle K_S|K_L\rangle  = (\epsilon +\epsilon^*)/ (1+|\epsilon|^2) \simeq 3.2 \times
10^{-3}$, we neglect this small $CP$--violation effect by taking  
$K_S$ and $K_L$ as orthogonal states, $\langle K_S|K_L\rangle  =0$. 

The situation then mimics perfectly that of the two--path interferometers
previously discussed and admits the same formalism. The approximation 
$\langle K_S|K_L\rangle =0$ just introduced corresponds to the (similarly approximate)
two--path orthogonality $\langle\psi_{I}|\psi_{II}\rangle =0$, and the states  
(\ref{statetn}) are particular cases of the state (\ref{state}) with: 
\[
\label{ab}
a = {1 \over \sqrt{1 + e^{-\Delta \Gamma \tau}}} , \; \; \;  
b = {1 \over \sqrt{1 + e^{+\Delta \Gamma \tau}}} , 
\; \; \; \phi = - \Delta m\, \tau \; \; {\rm or} \; \; \pi - \Delta m\, \tau . \nonumber
\]
Similarly, within our $CP$--conserving approximation, the
strictly orthogonal $\{K^0,\bar{K^0}\}$ basis is related to the
$\{K_S,K_L\}$ basis by:
\begin{equation}
\label{Sbasis}
|{K^0}\rangle  = {1 \over \sqrt 2} [|K_S\rangle  + |K_L\rangle ], \; \; \;
|\bar{K^0}\rangle = {1 \over \sqrt 2} [|K_S\rangle  - |K_L\rangle ] ,
\end {equation}
in close analogy with Eq.~(\ref{pm}). In this basis one has:
\begin{eqnarray}
\label{statetnS}
|K^0 (\tau)\rangle  &=& {1 \over \sqrt{2}} 
\left[ {1 + e^{-{1\over 2}\Delta \Gamma \tau} e^{-i \Delta m \tau}  
\over \sqrt{1 + e^{-\Delta \Gamma \tau}}} |K^0\rangle  +
       {1 - e^{-{1\over 2}\Delta \Gamma \tau} e^{-i \Delta m \tau}
\over \sqrt{1 + e^{-\Delta \Gamma \tau}}} |\bar{K^0}\rangle  \right] , \\ 
\label{statetnS2} 
|\bar{K^0} (\tau)\rangle  &=& {1 \over \sqrt{2}} 
\left[ {1 - e^{-{1\over 2}\Delta \Gamma \tau} e^{-i \Delta m \tau}  
\over \sqrt{1 + e^{-\Delta \Gamma \tau}}} |K^0\rangle  +
       {1 + e^{-{1\over 2}\Delta \Gamma \tau} e^{-i \Delta m \tau}
\over \sqrt{1 + e^{-\Delta \Gamma \tau}}} |\bar{K^0}\rangle  \right], 
\end{eqnarray}
thus mimicking Eq.~(\ref{statepm}). 
For the ``fringe visibility'' and the \emph{a priori} ``predictability'' one
now obtains the \emph{time--dependent} expressions:
\begin{eqnarray}
\label{VPK}
{\cal V}_{0} (\tau)&=&  
{2 \over \sqrt{2+e^{-\Delta \Gamma \tau} + e^{+\Delta \Gamma \tau}}} = 
{1 \over \cosh \left( {1 \over 2} \Delta \Gamma \tau\right) },  \\
\label{VPK2}
{\cal P} (\tau) &=& \left| {1 \over 1 + e^{-\Delta \Gamma \tau}} - 
{1 \over 1 + e^{+\Delta \Gamma \tau}} \right|
 = \tanh \left( {1 \over 2} \Delta \Gamma \tau\right),
\end {eqnarray} 
which obviously verify for all values of $\tau$ the time--dependent version 
of Eq.~(\ref{PV=}):
\begin{equation}
\label{PV=t}  
{\cal P}(\tau)^2 + {\cal V}_{0} (\tau)^2 =1 , 
\end{equation}
as expected for pure states such as $|{K^0} (\tau)\rangle $ and $|\bar{K^0} (\tau)\rangle $. 

The physical interpretation of these kaonic results seems unique and quite
obvious. As soon as a $K^0$ or a $\bar{K^0}$ is produced by
strangeness--conserving strong interactions, it starts propagating in free
space in the coherent superposition (\ref{Sbasis}) of the 
$|K_S\rangle$ and $|K_L\rangle$
components. These two components, which evolve in time without oscillating
into each other, decrease exponentially at remarkably different decay rates,
$\Gamma_S \simeq 579\, \Gamma_L$.  The $|K_S\rangle $ and $|K_L\rangle $ states are thus the
analogs of the two separated paths $|\psi_I\rangle $ and $|\psi_{II}\rangle $ 
(associated with particle--like behaviour) in usual interferometers, with 
$\langle K_L|K_S\rangle \simeq \langle\psi_{II}|\psi_I\rangle \simeq 0$. 
Much in the same way as the probabilities for taking each one of the two
paths are not equal in asymmetric interferometers, the
probabilities for $|K_S\rangle $ or $|K_L\rangle $ propagation are similarly different
except at $\tau=0$. Indeed, for undecayed kaons surviving up to $\tau>0$ one knows
that  (slowly decaying) $|K_L\rangle $ propagation is more likely than its  
(much faster decaying) alternative $|K_S\rangle $. One has an \emph{a priori} knowledge or 
``predictability'' on the actual propagation path, which is merely a consequence
of knowing the state one is dealing with, as in the case of asymmetric
interferometers. 

The role of ``fringe visibilty'' in ordinary interferometry 
(associated to the complementary wave--like behaviour) is played by the
well known phenomenon of strangeness oscillations in the neutral kaon case. 
As previously mentioned, in ordinary interferometers 
(such as those of Refs.~\cite{RS84,MPS87}) the two amplitudes
corresponding to the $|\psi_I\rangle $ and $|\psi_{II}\rangle $ paths have to be recombined
before emerging from the two output ports, where a
projective measurement in the $\{\psi_+,\psi_{-}\}$ basis is performed. 
In the neutral kaon case one takes advantage of the strong analogy
between the bases (\ref{pm}) and (\ref{Sbasis}), and 
of the fact that the
amplitudes for $|K_S\rangle $ and $|K_L\rangle $ propagation are
automatically recombined if a projective measurement 
in the $\{K^0,\bar{K^0}\}$ basis, corresponding to strangeness $S=\pm 1$, is performed.
There are two independent and time--honoured methods for
performing these measurements (a short historical review may be found
in Ref.~\cite{CPLEARstrong}) and both have been successfully used by the
CPLEAR collaboration, as we now proceed to discuss.

%\vspace{0.5cm}
%{\bf 4.- CPLEAR experiments}
\section{CPLEAR experiments}

In a recent CPLEAR experiment \cite{CPLEARstrong}, $K^0$--$\bar{K}^0$ oscillations
have been observed via strangeness measurements monitored
by kaon--nucleon strong interactions. The
previously mentioned proton--antiproton
annihilation processes, $p \bar p \to K^- \pi^+ K^0$ or $p \bar p \to K^+ \pi^-
\bar{K^0}$, were used 
to produce initial $|K^0\rangle $ or $|\bar{K^0}\rangle $ states, 
which were allowed to propagate in free space. 
The strangeness of the states $|K^0 (\tau)\rangle $ and $|\bar{K^0}(\tau)\rangle$, 
Eqs.~(\ref{statetnS}) and (\ref{statetnS2}), was
subsequently measured at different proper times $\tau$. This was achieved by
inserting a 2.5 cm thick carbon absorber which allowed to detect
neutral kaon interactions with nucleons in the time interval $1.3$--$5.3\, \tau_S$. 
The number of $K^0$'s and $\bar{K^0}$'s interacting with the
absorber's bound nucleons via $K^0 +p \to K^+ +n$ (thus projecting on 
the $|K^0\rangle $ state) or, alternatively, via $\bar{K^0} +n \to K^- +p$, 
$\bar{K^0} +n \to \pi^0 + \Lambda (\to \pi^- +p)$ (projecting on $|\bar{K}^0\rangle$)
were carefully recorded together with the vertex position or interaction time
$\tau$. An asymmetry parameter, $A_{\Delta m}^{\rm strong} (\tau)$, conveniently 
minimizing some experimental uncertainties, was then defined (see Eq.~(5) in
Ref.~\cite{CPLEARstrong}) and used to extract a value for
$\Delta m$ ---fully compatible with other measurements---
by fitting the time--dependence of these data. 

These  findings, however, admit an independent and immediate interpretation in
terms of the ``kaon interferometry'' we are discussing. Indeed, the  measured
asymmetry parameter, $A_{\Delta m}^{\rm strong} (\tau)$, 
can be easily rewritten as:
\[
A_{\Delta m}^{\rm strong} (\tau)= {2 {\cal V}_0(\tau) \cos (\Delta m \tau)
\over 1 + {\cal V}_0^2(\tau) \cos^2 (\Delta m \tau)} , \nonumber
\]
with ${\cal V}_0(\tau) = 1/\cosh \left(\Delta \Gamma \tau/2 \right)$, as
given in Eq.~(\ref{VPK}). In other words, the CPLEAR data \cite{CPLEARstrong} 
can be viewed as the successful measurement of the ``fringe visibility'' of
strangeness oscillations according to our discussion. By jointly considering
these ${\cal V}_0 (\tau)$ measurements with the complementary ``which path'' information
${\cal P}(\tau)$ ---which is simply given by inserting the observed interaction time
$\tau$ and the well known values of $\Gamma_S$ and $\Gamma_L$ in  
Eq.~(\ref{VPK2})---, these CPLEAR results are seen to fulfil the statement for
quantitative duality, Eq.~(\ref{PV=t}), for the whole range of $\tau$--values 
$1.3$--$5.3\, \tau_S$. 

In another CPLEAR experiment \cite{CPLEARlepton}, equivalent results were
obtained and the whole picture is confirmed. Here strangeness oscillations were
observed by detecting semileptonic neutral kaon decays. According to the
well tested $\Delta S = \Delta Q$ rule, the ${K^0} \to e^- \pi^+ \bar{\nu}$ 
and $\bar{K^0} \to e^+ \pi^- {\nu}$ decays are forbidden and, therefore,  
whenever observed, these semileptonic final states have to be 
necessarily associated with weak decays from $\bar{K^0}$ and ${K^0}$ states,
respectively. This represents an independent method of strangeness
measurement and a new asymmetry parameter,
$A_{\Delta m}^{\rm weak} (\tau)$  (Eq.~(2) in Ref.~\cite{CPLEARlepton}), can be
defined in such a way that Eq.~(\ref{I}) becomes 
$I(\tau)_{\pm}=[1\pm A_{\Delta m}^{\rm weak}(\tau)]/2$. 
The $A_{\Delta m}^{\rm weak} (\tau)$ measurement shows again a
characteristic oscillatory  $\tau$--dependence and has been fitted to extract an
independent value for $\Delta m$, in agreement with 
the present world average and even having the same accuracy.
Alternatively, one can reinterpret these results as before. 
Indeed, assuming no violation of the $\Delta S = \Delta Q$ rule, one easily
finds: 
\[
A_{\Delta m}^{\rm weak} (\tau)= {\cal V}_0(\tau) \cos (\Delta m \tau)=
{\cos (\Delta m \tau)
\over \cosh\left({1 \over 2}\Delta \Gamma \tau\right)} , \nonumber 
\]
as required by Eq.~(\ref{VPK}). Again, using these CPLEAR data to 
extract values of ${\cal V}_0 (\tau)$ and ${\cal P}(\tau)$ would confirm the validity of  
Eq.~(\ref{PV=t}) from $\tau \simeq  1.5\, \tau_S$ to $\tau \simeq 20\, \tau_S$.

Prompt strangeness measurement events at $\tau<<\tau_S$, for which one
predicts ${\cal V}_0 (\tau<<\tau_S)\simeq 1$ and ${\cal P} (\tau<<\tau_S)\simeq 0$,
were not performed by the CPLEAR collaboration. Note, however, that both sets of
CPLEAR data include measurements around $\tau=1.8\, \tau_S$ showing a
%Note that for prompt strangeness measurement events around $\tau \simeq 2 
%\tau_S$, one observes (in both sets of CPLEAR data) a well 
contrasted oscillatory behaviour [${\cal V}_0(\tau = 1.8\, \tau_S)\simeq 0.7$] 
because the available information on which component, $K_S$ or
$K_L$, is actually propagating is still incomplete 
[${\cal P}(\tau = 1.8\, \tau_S)\simeq 0.7$] and cannot be in any way
increased. Indeed, these observed neutral kaons have been converted into
another hadron or a semileptonic final state once their strangeness has been
measured as in Refs.~\cite{CPLEARstrong} and \cite{CPLEARlepton}, respectively.
Conversely, kaons surviving up to $\tau \simeq 5\, \tau_S$ or more are known to
propagate as $K_L$'s almost for sure [${\cal P}(\tau \gsim 5\, \tau_S)\simeq 1$].
Indeed, the probability that a $K_S$ survives up to $5\, \tau_S$ 
%and is then wrongly identified as a $K_L$ 
reduces to a few per thousand,
%\cite{bg1}, 
which is of the same order as the
$CP$--violation effects we are systematically neglecting, and thus beyond the
accuracy of our present approximate treatment. Consequently, the contrast of strangeness
oscillations is seen to disappear when approaching these larger $\tau$--values. 

%\vspace{0.5cm}
%{\bf 5.- Conclusions}
\section{Conclusions} 

In our view, the previous Section exemplifies in a clear way the concept of
quantitative duality, for which a general formulation has been recently
proposed by Englert in the following terms \cite{englert96}: 
\begin{quotation}
\emph{Duality}---The observation of an interference pattern and the
adquisition of which--way information are mutually exclusive.
\end{quotation}
As anticipated by Feynman's quotations in the Introduction, the
extension of these ideas to ``neutral kaon interferometry'' is extremely simple
and illustrative. The transition from maximal ``fringe visibility'' at $\tau=0$ ---with no
information on which component actually propagates--- to the opposite extreme
case, $\tau>>\tau_S$, allows one to cover the intermediate stages automatically by measuring
strangeness at intermediate times $\tau$. 
This leads to simple $\tau$--dependent expressions for ${\cal V}_0 (\tau)$ and ${\cal P}(\tau)$
in terms of $\Delta \Gamma\, \tau$, whereas the oscillatory term itself is exclusively 
driven by $\Delta m\, \tau$. In this sense, neutral kaon interference experiments
in free space are universally governed by these two well determined parameters alone:
$\Delta \Gamma$ and $\Delta m$. Note however that the time--dependence of 
${\cal V}_0 (\tau)$ and ${\cal P}(\tau)$, which cannot be avoided for unstable kaons, 
is specific of our context. It allows for further and independent tests
of basic principles in terms of  
Eq.~(\ref{PV=t})\footnote{Most frequently, experiments are expected to test Eq.~(\ref{PV=})
[rather than Eq.~(\ref{PV=t})] where ${\cal V}_0$ and ${\cal P}$ are constants
in a given experimental set--up. A curious exception is the optical interferometer
proposed in Ref.~\cite{bartell80}, where the ``fringe visibility'' is found to have
the same functional dependence (in space) as Eq.~(\ref{VPK}) (in time).}. 

To the best of our knowledge, only the neutron experiments of Ref.~\cite{RS84} and the
photon experiment of Ref.~\cite{MPS87} have attempted
to test these ideas. In both cases, an older version of the interferometric
duality--- which was originally introduced in Ref.~\cite{WZ79} and, according to
Ref.~\cite{JSV95}, involves a less satisfactory measure of the ``which--way''
information (the information--theoretical ``lack of knowledge'' rather than
``predictability'')--- was used. 

We have shown that two CPLEAR experiments,
which admit a transparent interpretation in terms of Eq.~(\ref{PV=t}), are fully
consistent with this equality. Further experiments at the operating
$\phi$--factory Daphne \cite{handbook}, which copiously produces neutral kaons via strong 
$\phi \to K^0 \bar{K^0}$ decays, are going to be of interest.

%\vspace{0.5cm}
%{\bf Acknowledgements} 
\section*{Acknowledgements}

This work has been partly supported by EURIDICE
HPRN--CT--2002--00311, BFM--2002--02588 and INFN.
Discussions with M.~Nowakowski are acknowledged.

%%%%%%%%%%%%%%%%%%%%%%%%%%%%%%%%%%%%%%%%%%

%\end{references}

\end{document}